\documentclass[twocolumn]{aastex63}
\usepackage{color, mhchem, here}
\usepackage{bm}
\received{}
\revised{}
\accepted{}
\submitjournal{ApJ}

\shorttitle{Various-type star photoevaporation}
\shortauthors{Komaki et al.}

\newcommand{\rg}{r_{\rm g}}
\newcommand{\kms}{{\rm km\,s^{-1}}}
\newcommand{\au}{{\rm \,au}}
\newcommand{\braket}[1]{\left(#1\right)}
\newcommand{\Msun}{{\rm \, M_\odot}}
\newcommand{\cm}{{\rm \, cm}}
\newcommand{\eV}{{\rm \, eV}}
\newcommand{\keV}{{\rm \, keV}}
\newcommand{\dd}{{\rm d}}
\newcommand{\Kelvin}{{\rm \,K}}
\newcommand{\Myr}{{\,\rm Myr}}
\newcommand{\yr}{{\,\rm yr}}
\newcommand{\figref}[1]{Figure~\ref{#1}}
\newcommand{\tabref}[1]{Table~\ref{#1}}
\newcommand{\eqnref}[1]{Eq.~\ref{#1}}
\newcommand{\secref}[1]{Section~\ref{#1}}

\begin{document}

\title{Radiation hydrodynamics simulations of protoplanetary disks: Stellar mass dependence of the disk photoevaporation rate}

\author[0000-0002-9995-5223]{Ayano Komaki}
\affiliation{Department of Physics, The University of Tokyo, 7-3-1 Hongo, Bunkyo, Tokyo 113-0033, Japan}
\email{ayano.komaki@phys.s.u-tokyo.ac.jp}

\author[0000-0002-1803-0203]{Riouhei Nakatani}
\affiliation{RIKEN Cluster for Pioneering Research, 2-1 Hirosawa, Wako-shi, Saitama 351-0198, Japan}

\author[0000-0001-7925-238X]{Naoki Yoshida}
\affiliation{Department of Physics, The University of
Tokyo, 7-3-1 Hongo, Bunkyo, Tokyo 113-0033, Japan}
\affiliation{Kavli Institute for the Physics and Mathematics of the Universe (WPI), UT Institute for Advanced Study, The University
of Tokyo, Kashiwa, Chiba 277-8583, Japan}
\affiliation{Research Center for the Early Universe (RESCEU), School of
Science, The University of Tokyo, 7-3-1 Hongo, Bunkyo, Tokyo 113-0033, Japan}



\begin{abstract}
Recent multiwavelength observations suggest that inner parts of protoplanetary disks (PPDs) have shorter lifetimes for heavier host stars. 
Since PPDs around high-mass stars are irradiated by strong ultraviolet radiation, photoevaporation may provide an
explanation for the observed trend.
We perform radiation hydrodynamics simulations of photoevaporation of PPDs for a wide range of host star mass of $M_* =0.5$--$7.0\Msun$. We derive disk mass-loss rate $\dot{M}$, which has strong stellar dependence as $\dot{M} \approx 7.30\times10^{-9}(M_{*}/\Msun)^{2}\Msun\yr^{-1}$. The absolute value of $\dot{M}$ scales with the adopted far-ultraviolet and X-ray luminosities. 
We derive the surface mass-loss rates and provide polynomial function fits to them. 
We also develop a semianalytic model that well reproduces the derived 
mass-loss rates.
The estimated inner-disk lifetime decreases 
as the host star mass increases, in agreement with the observational trend. 
We thus argue that 
photoevaporation is a major physical mechanism for PPD dispersal for a wide range of the stellar mass and can account for the observed stellar mass dependence of the inner-disk lifetime. 

\end{abstract}



\section{Introduction} \label{sec:intro}
Recent infrared observations toward nearby star-forming regions revealed that the fractions of stars with circumstellar disks (protoplanetary disks; PPDs) decline with increasing the age of the star-forming region. 
The disk fraction as a function of age gives an approximate measure of the disk lifetime, and the results of the recent observations consistently suggest $t_{\rm life} = 3$--$6 \Myr$ \citep[e.g.,][]{Haisch2001, Meyer:2007, Hernandez:2007, Mamajek:2009, 2012Bayo, Ribas:2014}.
Planets are formed from the gas and dust grains in PPDs, where dynamical interaction between planets and the parent disk shapes the initial configuration of the planetary system.
The disk lifetime is thus an important factor in the study of planetary system formation.

There has been an intriguing suggestion from systematic infrared observations that the disk lifetime depends on the host star mass. Disks around massive stars have shorter lifetimes than around low-mass stars
\citep[e.g.,][]{Carpenter:2006,Lada:2006, Allers:2007, DahmHillenbrand:2007, KennedyKenyon:2009, Fang:2012, Yasui:2014, Ribas:2015}. 
\cite{Ribas:2015} observe a number of associations and 
conclude that the disk lifetimes are systematically shorter for systems with a central star mass of $>2 \Msun$ than those with $< 2\Msun$.
Similarly, \citet{2012Bayo} show that the disk fraction monotonically decreases as the mass of the central star increases in the range $0.1$--$1.7\Msun$.
An apparently opposite trend has also been reported. 
Observations using Atacama Large Millimeter/submillimeter Array (ALMA) suggest that the {\it dust}-disk mass decreases
systematically faster around lower-mass stars with $\sim 0.1$--$3\Msun$ \citep{Ansdell:2016, Pascucci:2016, 2017Ansdell}. 
The physical mechanism that can account for these observations remains unknown.

Theoretical studies suggest three major disk-dispersal mechanisms: accretion \citep{Shakura:1973, Lindenbell:1974}, magnetohydrodynamic (MHD) winds \citep[e.g.,][for a review]{Suzuki:2009, Turner:2014}, and photoevaporation \citep[e.g.,][for a review]{1994_Hollenbach,1994_Shu, Alexander:2014}. Disk materials fall onto the host star by losing the angular momentum through viscous friction, while being stripped off from the disk surface by the winds. 
The disk dynamics and the mass evolution are determined by these effects mutually affecting each other \citep[e.g.,][]{Clarke:2001, Alexander:2006, Owen:2010, Wang:2019, Gressel:2020}. 
The strengths of the effects and the dispersal timescale depend on the properties of the disk and the host star.

Photoevaporation is driven by the stellar far-ultraviolet (FUV; $6\eV \lesssim h\nu < 13.6\eV $), extreme-ultraviolet (EUV; $13.6\eV < h\nu \lesssim 100\eV$), and X-rays ($100\eV \lesssim h\nu \lesssim 10\keV$). Thermalizing electrons produced by photochemical reactions heat the disk gas, and the hot gas flows out of the disk and escapes from the gravitational potential of the central star. 
EUV photons are primarily absorbed by atomic hydrogen, while FUV and X-ray are absorbed also by dust grains and gas-phase metal species. 
FUV and X-ray photons can penetrate into deep interior regions of a disk with high column densities ($\sim 10^{21}\cm^{-2}$)
\citep[e.g.,][]{RichlingYorke:2000, Ercolano:2009, GortiHollenbach:2009} 
and thus drive high-density photoevaporative flows. 
The rate of photoevaporation $\dot{M}$ depends on the stellar UV and X-ray luminosities, and it is expected that $\dot{M}$ increases with increasing stellar mass.
In the light of this, the exact dependence of $\dot{M}$ on
the stellar mass can provide a clue to understanding the physical mechanism of disk dispersal and the observed trend of the disk lifetime.


In the present paper, we perform a set of
radiation hydrodynamics simulations of disk photoevaporation.
In order to examine the stellar mass dependence quantitatively, we systematically vary the host star mass
and derive the disk photoevaporation rate when the system
reaches a quasi-steady state.
\cite{GortiHollenbach:2009} explore the stellar mass dependence by studying the thermochemical structure of disks in hydrostatic equilibrium. They obtain $\dot{M}$ 
for stars masses with $0.3$--$7.0\Msun$.
Ideally, fully coupled radiation hydrodynamics simulations including thermochemistry are essential to derive the disk structure and the resulting $\dot{M}$ self-consistently \citep{WangGoodman:2017, Nakatani:2018a, Nakatani:2018b}.
It is also important to follow the long-term evolution
of photoevaporating disks \citep{Nakatani:2018a}.  
We derive the stellar mass dependence, for the first time,
by using multidimensional radiation hydrodynamics simulations.

The rest of the present paper is organized as follows.
In Section 2, we explain the methods of our simulations.
In Section 3, we show the result of our simulations.
In Section 4, we discuss the result.
Finally, in section 5 we summarize.

\section{Numerical simulations}
We perform simulations of disk photoevaporation with varying the stellar mass in a wide range of 0.5--7.0$\Msun$.
The basic method is the same as \cite{Nakatani:2018a, Nakatani:2018b}. 
Briefly, we solve hydrodynamics, radiative transfer, and nonequilibrium chemistry in a fully coupled, self-consistent manner. 
Our simulations follow transfer of FUV, EUV, and X-ray radiation from the central star. The respective luminosities are adopted from Table~1 of \cite{GortiHollenbach:2009}
for stellar parameters of pre-main-sequence stars of age $\sim 1\Myr$ with $0.3$--$7\Msun$. The EUV and X-ray are assumed to originate from the chromosphere, while the FUV generated by accretion is additionally incorporated.
The fiducial stellar parameters we use are shown in \tabref{table:stellarparameter},
where $\phi_{\rm{EUV}}$, $L_{\rm{FUV}}$, and $L_{\rm{X-ray}}$ are the emission rate of EUV photons, FUV luminosity, and X-ray luminosity, respectively.
Since they are not well characterized by observations and 
can vary as the system evolves, we treat them as parameters 
and examine the overall effect on the disk mass-loss rate. In \secref{sec:atsomepoint}, we discuss the variability of photoevaporation rates by performing additional simulations with various luminosities for a few selected systems.
We assume that the FUV radiation is contributed by continuum emission, and we do not explicitly take into account the hydrogen Ly$\alpha$ line, which has been suggested to be a substantial contributor to the total FUV flux in systems of classical T Tauri stars \citep{2003:Bergin, 2004:Herczeg, 2012:Schindhelm}. In \secref{sec:atsomepoint} we show the results of additional simulations with higher $L_{\rm FUV}$. The simulations roughly correspond to the cases where the Ly$\alpha$ contribution is incorporated.

The disk gas contains eight chemical species; \ion{H}{1}, \ion{H}{2}, \ce{H2}, \ce{H2+}, \ce{CO}, \ion{O}{1}, \ion{C}{2} and electrons.
We assume that \ion{C}{1} is ionized to \ion{C}{2} immediately after photodissociation of \ce{CO} \citep{NelsonLanger:1997, RichlingYorke:2000}, so we do not include \ion{C}{1} in the simulations.
We set the dust-to-gas mass ratio to $0.01$, the gas-phase elemental abundances of carbon to $y_{\ce{C}}=0.927\times 10^{-4}$, and the gas-phase elemental abundance of oxygen to $y_{\ce{O}}=3.568\times 10^{-4}$ \citep{Pollack1994}.

\begin{table}
  \caption{Fiducial stellar parameters in the model (adopted from \citet{GortiHollenbach:2009})}
  \centering
  \begin{tabular}{cccc} \hline
    $M_*$ ($\Msun$) 
    & log$\phi_{\rm{EUV}}$ (/s) & log$L_{\rm{FUV}}$ (erg/s) & log$L_{\rm{X-ray}}$ (erg/s)\\ \hline 
    0.5 
    & 40.1 & 30.9 & 29.8\\
    0.7 
    & 40.5 & 31.3 & 30.2\\
    1.0 
    & 40.7 & 31.7 & 30.4\\
    1.7 
    & 41.0 & 32.3 & 30.7\\
    3.0 
    & 39.0 & 32.9 & 28.7\\
    7.0 
    & 44.1 & 36.5 & 30.8\\ \hline
  \end{tabular}
  \label{table:stellarparameter}
\end{table}


We assume an axisymmetric disk around the rotational axis and also assume symmetry with respect to the midplane.
The simulations are performed in 2D~spherical coordinates
with all the three components of velocity $\bm{v}=(v_{r}, v_{\theta}, v_{\phi})$.
The governing equations are 
\begin{gather}
\frac{\partial \rho}{\partial t} + \bm{\nabla} \cdot (\rho \bm{v}) = 0,\\
\frac{\partial (\rho v_{r})}{\partial t} + \nabla\cdot (\rho v_{r} \bm{v}) = - \frac{\partial P}{\partial r} - \rho\frac{GM}{r^2} + \rho\frac{v_{\theta}^2+v_{\phi}^2}{r},\\
\frac{\partial (\rho v_{\theta})}{\partial t} + \nabla\cdot (\rho v_{\theta} \bm{v}) = - \frac{1}{r}\frac{\partial P}{\partial \theta} - \rho\frac{v_{r}v_{\theta}}{r} + \rho\frac{v_{\phi}^2}{r}\cot\theta,\\
\frac{\partial (\rho v_{\phi})}{\partial t} + \nabla^{l}\cdot(\rho v_{\phi}\bm{v}) = 0,\\
\frac{\partial E}{\partial t} + \nabla \cdot H \bm{v}
= - \rho v_{r}\frac{GM}{r^2} + \rho(\Gamma - \Lambda),\\
\frac{\partial n_{\ce{H}}y_{i}}{\partial t} + \nabla\cdot (n_{\ce{H}}y_{i}\bm{v}) = n_{\ce{H}}R_{i}.
\end{gather}
The first five equations are for fluid equations, and the last one means the conservation of chemical species.
In the equations, $\rho$, $\bm{v}$, $P$, and $M_{*}$ represent gas density, velocity, pressure, and stellar mass, respectively.
The fifth equation is the energy equation.
$E$ and $H$ are the total energy and enthalpy including kinetic energy per unit volume.
$\Gamma$ and $\Lambda$ are the specific heating rate and specific cooling rate, which means the heating and cooling rate per unit mass.
The azimuthal component of the Euler equation is described in the angular momentum conservation form. 
We define $y_{i}$ as an abundance of chemical species $i$,
$n_{\rm H}$ as the number density of elemental hydrogen, 
and $R_{i}$ as a chemical reaction rate.
We take into account relevant chemical reactions tabulated in \citet{Nakatani:2018a, Nakatani:2018b}.
We use PLUTO \citep{2007_Mignone} to solve hydrodynamics.

We include EUV/X-ray photoionization heating, photoelectric heating \citep{1994BakesTielens}, dust-gas collisional cooling \citep{YorkeWelz:1996}, fine-structure cooling of \ion{C}{2} and \ion{O}{1} \citep{1989HollenbachMcKee, 1989Osterbrock, 2006Santoro}), molecular line cooling of $\ce{H2}$ and $\ce{CO}$ \citep{1998Galli, 2010Omukai}, hydrogen Lyman $\alpha$ line cooling \citep{1997Anninos}, and radiative recombination cooling \citep{1978Spitzer}.

Dust temperatures are determined in the same manner as in \cite{Nakatani:2018a, Nakatani:2018b} by performing 2D~radiative transfer calculations with a hybrid scheme \citep{2010Kuiper, 2013KuiperKlessen}. Both the direct stellar irradiation and dust re-emission are taken into account. We use the dust opacity of \cite{1984Draine} that assumes an interstellar dust model.

By running a number of radiation transfer calculations, we have found that the dust temperature distribution is well 
described by a function of radial distance, $R$, and column density, $N_{\rm H}$.
The average relative errors are 
$14\%$ for $M_{*}=0.5$--$3.0\Msun$
and $24\%$ for $M_{*}=7.0\Msun$,
and the error gets relatively large ($\gtrsim 80\%$) at large $R$ where $N_{\rm H}$ is small. However, in such outer low-density regions, the dust temperature is unimportant to determine the gas temperature. Therefore, the error or the inaccuracy of our fitting hardly affects our main results.  
We use a fitting function for local dust temperatures for each case with different stellar mass ($M_* = 0.5$--$7.0\Msun$).
This significantly saves the computational costs, and thus
we are able to follow the long-term evolution of PPDs.

We set the polar angle extent of the computational domains
to $0 \leq \theta \leq {\pi}/{2}$ rad.
We set the radial extent of the computational domains
to $0.1\rg \leq r \leq 20 \rg$,
where $\rg$ is the gravitational radius for ionized gas
defined by
\[
	\rg = \frac{GM_*}{(10\kms)^2} \simeq 8.87 \au \braket{\frac{M_*}{1\Msun}}.
\]
The radial extent is determined to cover the whole region where photoevaporative flows are expected to be driven \citep{2003Liffman}, and to avoid spurious reflection of subsonic flows at the outer boundary \citep{Nakatani:2018a}.

In agreement with viscous disk models \citep{Shakura:1973, Lindenbell:1974}, the mass accretion rates of low/intermediate-mass stars have been observed to decrease with increasing the age and with decreasing disk mass, $M_{\rm disk}$ \citep[e.g.,][]{Najita:2007, Mendigutia:2012}. 
For the initial condition, we assume evolved systems ($t \gtrsim 10^5$--$10^6\yr$), where the main accretion phase has ended and the accretion rate has been dropped (e.g., $\lesssim 1\times10^{-8}\Msun\yr^{-1}$ for $M_* \leq 1\Msun$). In such systems, photoevaporation (or MHD winds) dominates the disk mass loss \citep{Clarke:2001, Alexander:2006, Owen:2010, Suzuki:2016, Kunitomo:2020}. 
We set the initial disk mass to $M_{\rm disk}/M_* = 0.03$ \citep{AndrewsWilliams:2005}.
The surface density is assumed to scale as $\propto R^{-1}$ \citep{Andrews:2009}.
The disk mass within $R_{\rm in} \leq R \leq R_{\rm out}$ is
\[
	M_{\rm disk} = \int_{R_{\rm in}}^{R_{\rm out}} 2 \pi R \Sigma(R) \, \dd R
			= 2\pi \rg \Sigma_{\rm g} (R_{\rm out} - R_{\rm in}),
\]
where $\Sigma_{\rm g}$ is the surface density at the gravitational radius, $\Sigma(R) = \Sigma_{\rm g} (R/\rg)^{-1}$.
The inner ($R_{\rm in}$) and outer ($R_{\rm out}$) edges of a PPD differ significantly even for systems with similar stellar mass, and likely depend on the formation process, the system's age, and other physical parameters \citep[e.g.,][]{Ansdell:2018, Najita:2018, Johnston:2020}. 
For the purposes of the present study, we make a practical choice by setting $R_{\rm in}$ 
at the inner boundary of the computational domain ($0.1 \rg$). Since the total disk mass and photoevaporation rates are largely contributed by the outer region, using a smaller $R_{\rm in}$ does not affect these quantities. As for $R_{\rm out}$, we scale it with $\rg$ to treat ``weak-gravity'' regions ($R > \rg$) impartially for various stellar mass and set to be equal to the outer boundary of the computational domain ($20\rg$). 
In this case, $\Sigma_g$ is computed as
\[
	\Sigma_g \simeq \frac{M_{\rm disk}}{40\pi \rg^2}\\
		 \simeq 27.1{\rm \,g\,cm^{-2}} \braket{\frac{M_*}{1\Msun}}^{-1} .
\]
The initial gas temperature is given by $T_{\rm ini} = 100\Kelvin (R/0.1\rg)^{-1/2}$. We note that the simulation results do not strongly depend on $T_{\rm ini}$ because of
rapid gas-dust collisional coupling in optically thick regions.

We run the simulations over a long time of $8.4\times 10^{3}\left( M_{*}/\Msun \right) \yr$, which is 10 times the crossing time of flows having velocities of $1\kms$ ($= 20\rg/1\kms \approx 8.4\times10^{2}\left(M_{*}/\Msun \right)\yr$). 
It is sufficiently long so that the time average of the mass-loss rate converges.

\section{Results}   \label{sec:results}
\subsection{Structure of Photoevaporating Disks}\label{sec:result1}
\begin{figure*}[htbp]
       \centering
         \includegraphics[width=\linewidth,clip]{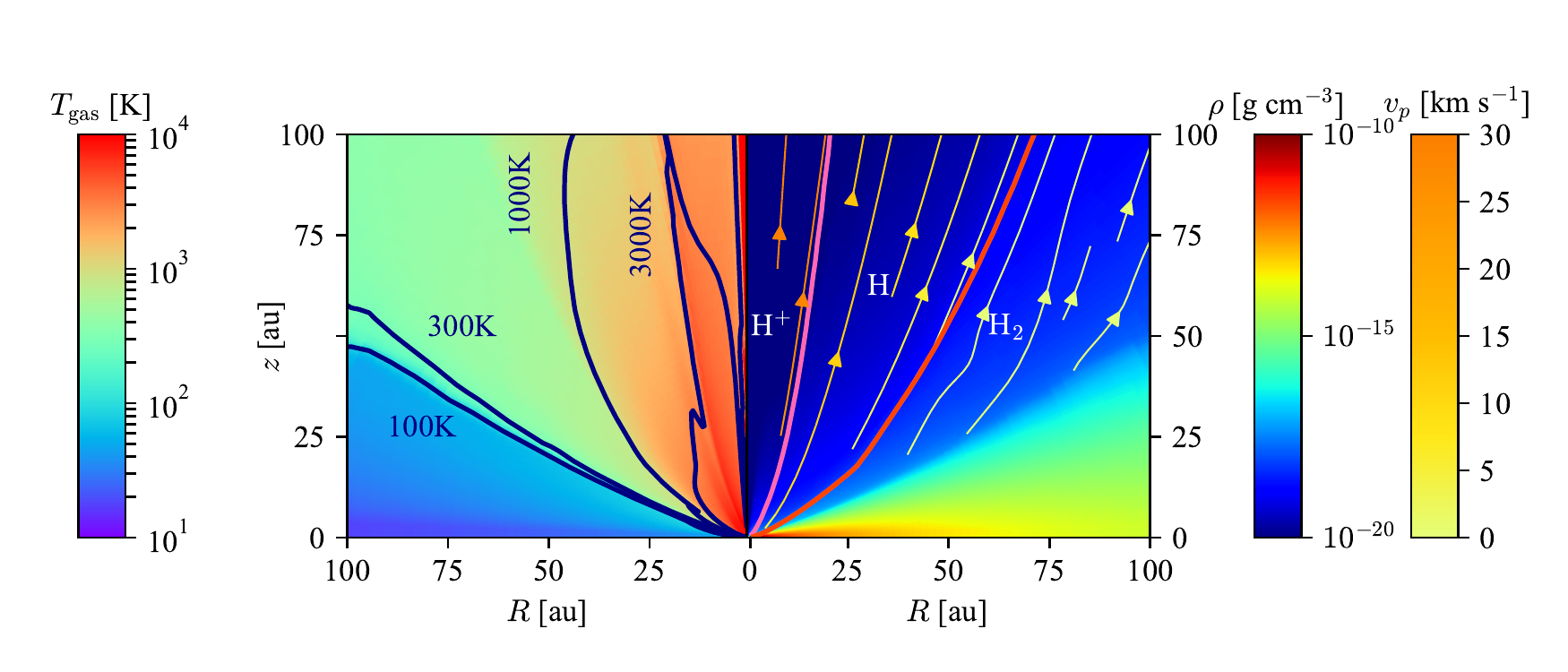}
         \caption{Disk structure at $t = 3000\yr$ for the run with $M_{*}=1.0\Msun$. 
         The color map on the left shows the gas temperature, and the right side shows the density distribution. The navy contour lines on the left show where the gas temperature is 100\,K, 300\,K, 1000\,K, and 3000\,K. The right portion also shows the chemical structure of H-bearing species. The pink line on the right side represents the ionization front where the abundance of \ion{H}{2} is 0.5, and the red line indicates the dissociation front where the abundance of $\ce{H2}$ is 0.25. The streamline represents the poloidal velocity of the gas.
         $v_{p}$ expresses the poloidal velocity.
         For clarity, we omit to plot the streamlines where the velocity is less than $0.1\kms$, which are mostly seen in the optically thick disk region. 
         }
      \label{fig:snapshots}
\end{figure*}

\figref{fig:snapshots} shows the snapshot at $t=3000\yr$ for the run with $M_* =1.0\Msun$.

In the right portion, the green region expresses the disk and the blue region is the evaporating gas.
Photoevaporative flows come from the surface of the disk where the column density of \ce{H2} satisfies $N(\ce{H2}) = 10^{20} \rm{cm}^{-2}$. 
We empirically define this boundary as the launching ``base'' of photoevaporative flows.
The temperature at the base depends on the central stellar mass (luminosity), which controls the mass-loss rate.
EUV photons heat the gas near the center where $y_{\ce{H\sc{II}}} \geq 0.5$ and the gas temperature 
$T \geq$ 3000K.
Adiabatic cooling is the dominant cooling process there.

\begin{figure}
         \includegraphics[width=\linewidth,clip]{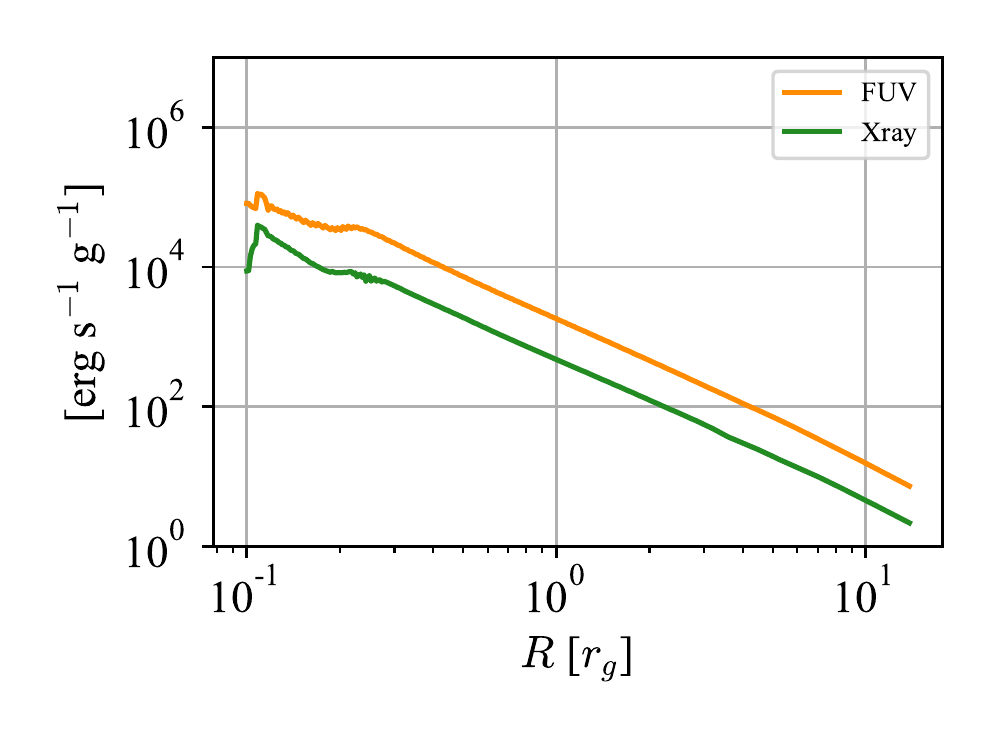}
         \caption{Volume heating rates of FUV photoelectric heating (orange) and the X-ray ionization heating (green) in the model with $M_{*}=1.0\Msun$. The distance $R$ is in units of $\rg$. 
         }
      \label{fig:heating}
\end{figure}

FUV and X-ray photons heat the gas in the disk interior and drive outflow of a neutral gas.
We define $\Gamma_{\rm{EUV}}$, $\Gamma_{\rm{FUV}}$ and $\Gamma_{\rm{X-ray}}$ as the heating rates by EUV, FUV and X-ray photons, respectively.
\figref{fig:heating} shows $\Gamma_{\rm{FUV}}$ and $\Gamma_{\rm{X-ray}}$ at the base.
Both the FUV and X-ray heating rates decrease toward the outer part. FUV heating is stronger than X-ray at all distances in the computational domain. This is also the case for simulations with other stellar masses.
The ratio of FUV to X-ray heating $\Gamma_{\rm{FUV}}/\Gamma_{\rm{X-ray}}$ ranges from 2 to 10 for $M_{*}=0.5$--$1.7\Msun$, and the ratio exceeds 1500 for the runs with $M_{*}=3.0$ and $7.0\Msun$.
The large heating ratio results from our setting that 
the increase of X-ray luminosity is smaller than that of FUV luminosity with increasing the stellar mass (see Table~\ref{table:stellarparameter}).
Note that the intermediate-mass stars do not have convective zones. The dynamo process of a $M_{*}=3.0$--$10\Msun$ star stops at the age of $\sim 1 \Myr$ \citep{2003Flaccomio},
and the weak magnetic activity on the stellar surface results
in X-ray radiation emission.
The FUV luminosity can vary according to the accretion rate onto the star. In \secref{sec:atsomepoint}, We discuss variations of the FUV/X-ray heating rates and the resulting mass-loss rates by running additional simulations with a wide range of stellar luminosities.

\cite{GortiHollenbach:2009} argue that FUV heating is the major driver of photoevaporation for PPDs around stars with $M_{*}=0.3$--$7.0\Msun$.
The temperature at the outflow launching ``base'' decreases with distance from the central star. 
We find $\Gamma_{\rm{FUV}}\propto (R/\rg)^{-2.0}$ at the base 
from our simulation results.
The X-ray heating rate has similar dependence on the distance from the central star to FUV.

\subsection{Photoevaporation Rate}
\begin{figure}
         \includegraphics[width=\linewidth,clip]{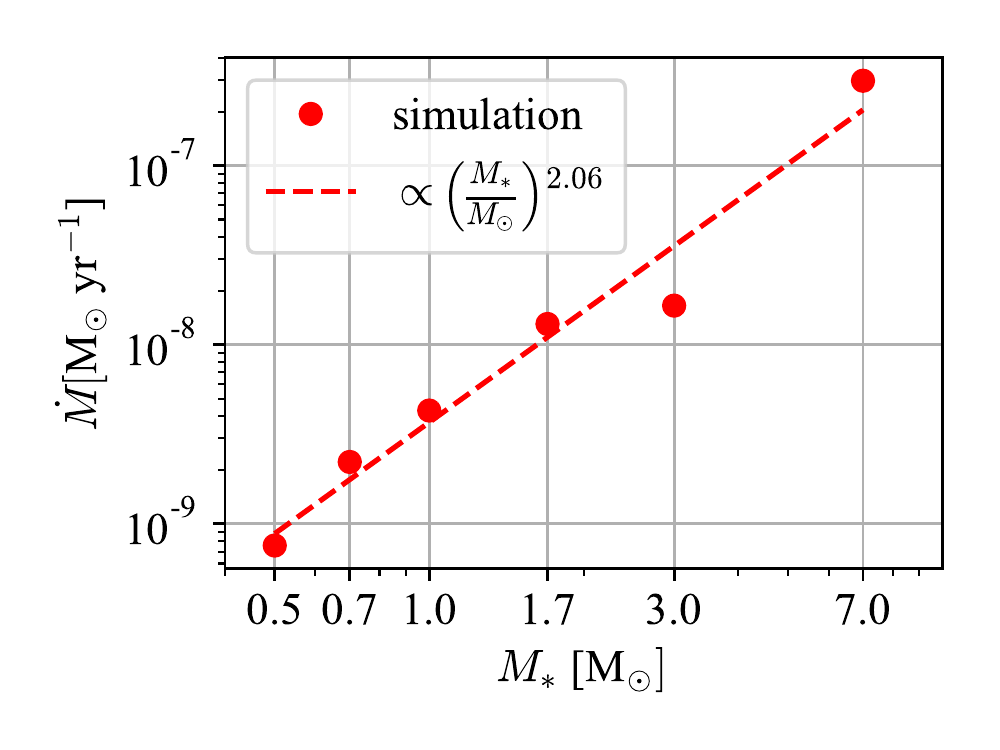}
         \includegraphics[width=\linewidth,clip]{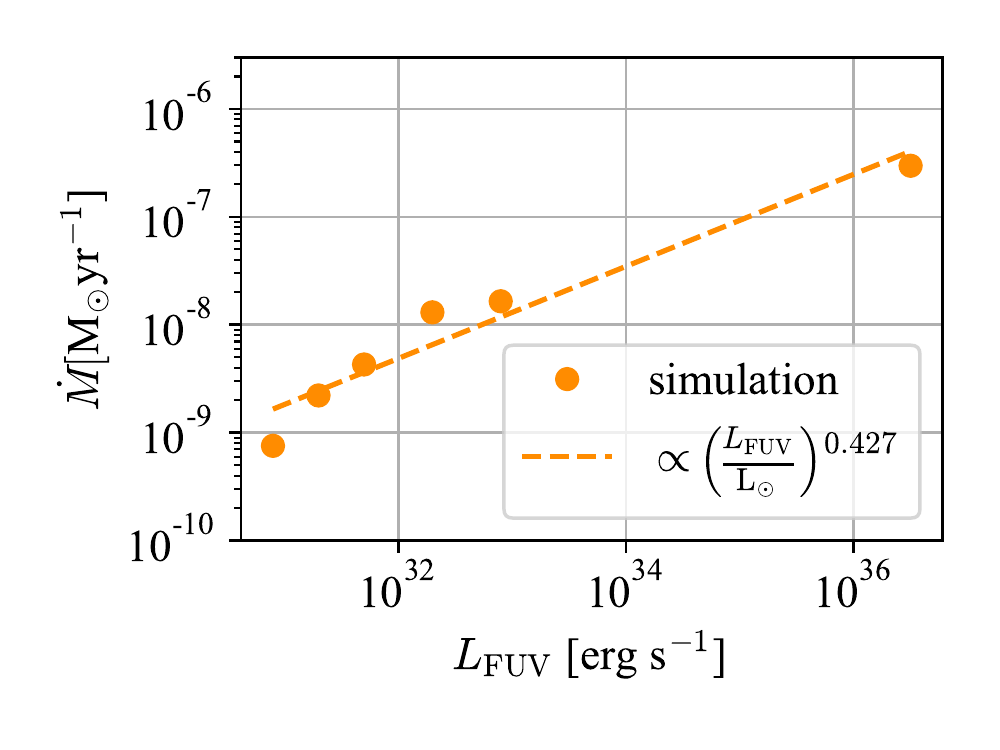}
         \caption{(top) Mass-loss rate with each mass of the central star. The red circles are the simulation data. The red dashed line is a fit. 
         (bottom) Mass-loss rates as a function of FUV luminosity. 
         The orange circles are the simulation data. 
        The lower $L_{\rm FUV} $ corresponds to the lower stellar mass (see \tabref{table:stellarparameter}). 
        The orange dotted line is a fit. 
         }
      \label{fig:simulationfuvmassloss}
\end{figure}

We calculate the mass-loss rate directly from our simulation outputs as
\begin{equation}
\dot{M} = \int_{S, \eta >0} \rho \bm{v}\cdot d\bm{S}, \label{eq:massloss}
\end{equation}
where $d\bm{S}$ is a spherical surface with radius of $20\rg$.
We define $\eta$ as specific enthalpy
\begin{equation}
\eta = \frac{1}{2}v_{p}^{2}+\frac{1}{2}v_{\phi}^{2}+\frac{\gamma}{\gamma -1}c_{s}^{2} - \frac{GM_{*}}{r}. \label{eq:enthalpy}
\end{equation}
The photoheated gas gains more energy than potential energy and thus evaporates if $\eta > 0$. We include the gas with $\eta > 0$ to calculate the mass-loss rate. The time-averaged mass-loss rates are shown in \figref{fig:simulationfuvmassloss}.

The disk mass-loss rate increases with the central star's mass;
the relationship is approximated as
\begin{equation}
\dot{M} (r < 20r_{\rm g}) \simeq 3.69\times 10^{-9}\left(\frac{M_{*}}{\Msun}\right)^{2.06}  \Msun\yr^{-1}
\label{eq:masslossslope}
\end{equation}
The gas temperature at $r=1\rg$ on the disk surface is $\simeq 10^3$ K for the run with $M_* = 1 \Msun$ and that around a $7 \Msun$ star is $\simeq 8\times10^3$ K.
The average temperature of the gas is higher for heavier central stars.


We set our calculation domain to $0.1\rg \leq r \leq 20\rg$, where $\rg$ is proportional to the central stellar mass.
We note that not only the luminosities but also the adopted $R_{\rm out} (=20r_{\rm g})$ contribute to the stellar mass dependence of the photoevaporation rate in our model. 
In order to distinguish the effects of $R_{\rm out}$ on the mass-loss rate, we calculate the mass-loss rates by setting the radius of the spherical surface $S$ to $80\au$ for all of the cases. 
We find that
\[
\dot{M} (r<80\au ) \simeq 1.91\times 10^{-9}\left(\frac{M_{*}}{\Msun}\right)^{1.56} \Msun\yr^{-1}
\]
gives a reasonably accurate fit. 
The $M_*$ dependence is weak compared to \eqnref{eq:masslossslope}. Nevertheless, the total photoevaporation rates are significantly higher for heavier stars owing to the stronger UV (X-ray) radiation. 
Previous studies also suggest an apparent increase of photoevaporation rate with the disk radius or measuring radius \citep{Tanaka:2013, Nakatani:2018a, Wolfer:2019}.

\cite{GortiHollenbach:2009} suggest that the typical mass-loss rate for a disk around a $1.0\Msun$ star is $3.0\times 10^{-8}\Msun$/yr,
and this is about 3 times larger than our result.
The mass-loss rates are consistent with the results of previous studies despite different computational methods. \citet{GortiHollenbach:2009} estimate mass-loss rates using a hydrostatic model, while we measure those by performing hydrodynamics simulations. 
It has also been shown that the X-ray photoevaporation rates differ only by a small factor between hydrostatic and hydrodynamics models \citep{Ercolano:2009, Owen:2010}.



\begin{figure}
         \includegraphics[width=\linewidth,clip]{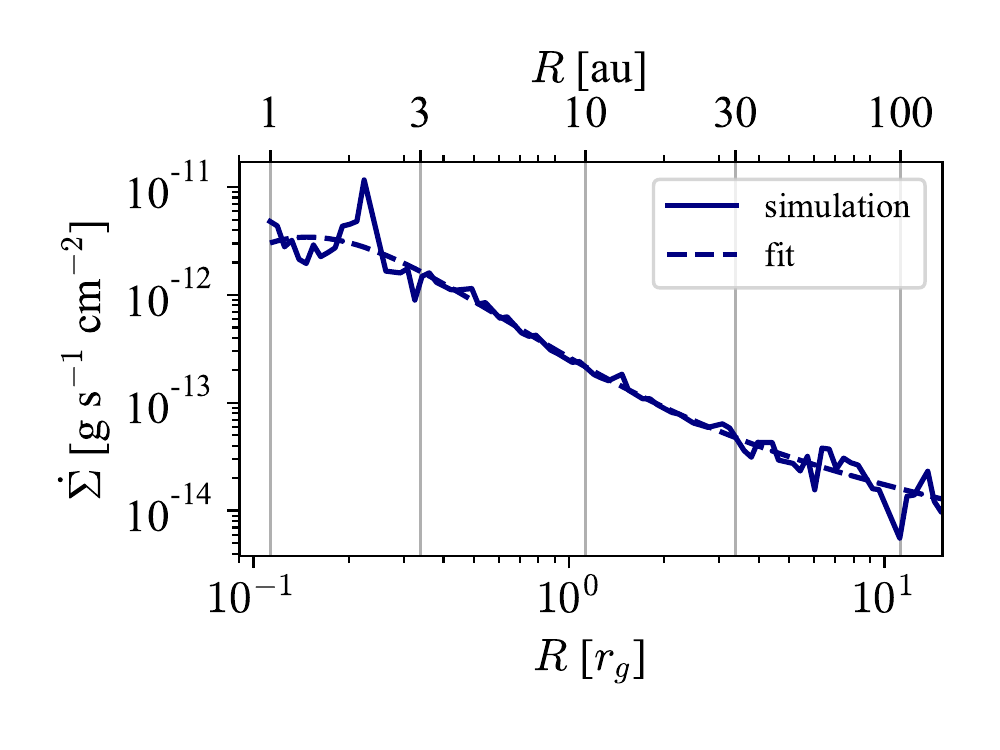}
         \includegraphics[width=\linewidth,clip]{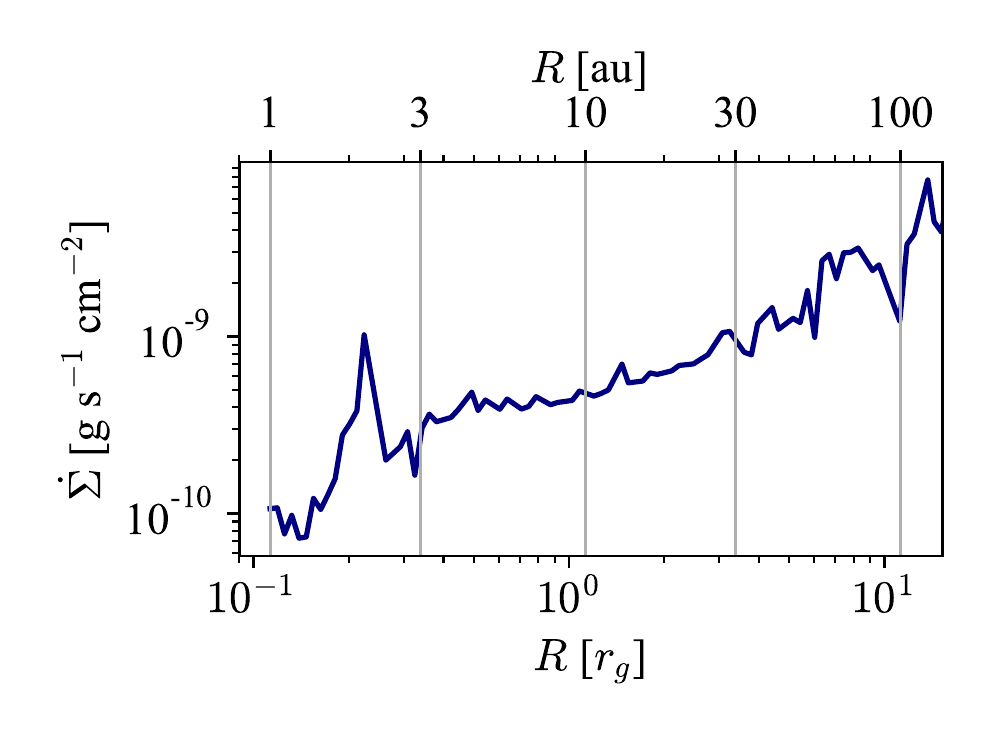}
         \caption{(top) Surface mass-loss rate $\dot{\Sigma}$ for our run with $M_{*}=1.0\Msun$. The horizontal axes show the radial distance in units of au and $\rg$ at the top and bottom, respectively. The surface mass-loss rate has a peak at $\sim 0.17\rg$ and decreases as the distance from the central star becomes large. The dotted line represents a fit given by equation \ref{eq:surfacemasslossrate}.
         (bottom) Weighted mass-loss rate $2\pi R^{2}\dot{\Sigma}(R)$ with $M_{*}=1.0\Msun$. The horizontal axes are the same as the top panel.}
      \label{fig:surfacedensity}
\end{figure}

We derive a time-averaged surface mass-loss rate, $\dot{\Sigma}(R)$, by differentiating the mass-loss rates with respect to the radius of the spherical surface $S$. Under steady state, the derivatives correspond to $2\pi R \dot{\Sigma}(R)$ by suitably transforming the spherical radius $r$ to the radial distance $R$.
We show the surface mass-loss rate, $\dot{\Sigma}(R)$, for the simulation with $M_* = 1.0\Msun$ in \figref{fig:surfacedensity}.
This figure clarifies which part of the disk evaporates most effectively.
The surface mass-loss rate at $M_{*}=1.0\Msun$ is approximated as
\begin{equation}
\begin{split}
\log_{10} \braket{\frac{\dot{\Sigma}(R)}{1\,{\rm g\,s^{-1}\,cm^{-2}}}} &= 0.380 x ^5-0.869x^4+0.303x^3\\
&+0.658x^2-1.67x-12.6,\\
\end{split}
\label{eq:surfacemasslossrate}
\end{equation}
where
\[
x  = \log_{10}\braket{\frac{R}{\rg}}.
\]
We approximate $\dot{\Sigma}(R)$ of other stellar masses in the same manner as above
\[
\begin{split}
\log_{10} \braket{\frac{\dot{\Sigma}(R)}{1\,{\rm g\,s^{-1}\,cm^{-2}}}} &= c_5\,x^5+c_4\,x^4 +c_3\,x^3\\
&+c_2\,x^2 +c_1\,x+c_0
\end{split}
\]
We obtain the fitting coefficients $c_0,c_1,c_2,c_3,c_4$, and $c_5$ as in \tabref{tab:fit}.
\begin{table}
  \caption{The coefficients of $\dot{\Sigma}(R)$}
  \label{tab:fit}
  \centering
  \begin{tabular}{L|C C C C C C}\hline
    M_* & c_5 & c_4 & c_3 & c_2 & c_1 & c_0 \\ \hline
    0.5 & 1.06 & -1.05 & -0.236 & 0.570 & -1.62 & -12.7 \\
    0.7 & 0.693 & -0.95 & -0.038 & 0.678 & -1.67 & -12.6 \\
    1.0 & 0.131 & -0.465 & 0.451 & 0.376 & -1.67 & -12.6 \\
    1.7 & 1.37 & -1.41 & -1.42 & 1.30 & -1.06 & -12.6 \\
    3.0 & 0.033& -0.786 & 0.786 & 0.557 & -1.58 & -13.1 \\
    7.0 & 0.594 & -1.00 & 0.234 & 0.513 & -1.85 & -12.1 \\ \hline
  \end{tabular}
\end{table}

We define $r_{\rm peak}$ as the radius at which $\dot{\Sigma}$ has the maximum value. 
In our run with $M_{*}=1.0\Msun$, the $\dot{\Sigma}$
peaks at $R=0.17\rg\sim1.5$ au. The peak position $r_{\rm peak}$ is larger for larger central stellar mass.
The peak marks a "boundary" that divides the photoevaporative flows 
into an FUV-driven region and an EUV-driven region; the flows are EUV driven inside the peak whereas they are FUV driven outside the peak. The inner, EUV-driven flows have smaller densities than FUV-driven flows, and thus the boundary appears as a peak in the mass-loss profile. 
We find nonzero mass flux in $R\leq 0.17\rg$. The gravitational binding by the central star is strong in this region, but photoevaporative flows can be launched initially subsonically, which are then accelerated to become supersonic flows as 
described by a Bernoulli flow model of \citet{Dullemond:2007}.

In order to see the effects of the choice of inner boundary size on $r_{\rm peak}$, we run additional simulations setting the computational domain to $0.03\rg\leq r\leq 20\rg$.
\begin{figure}
         \includegraphics[width=\linewidth,clip]{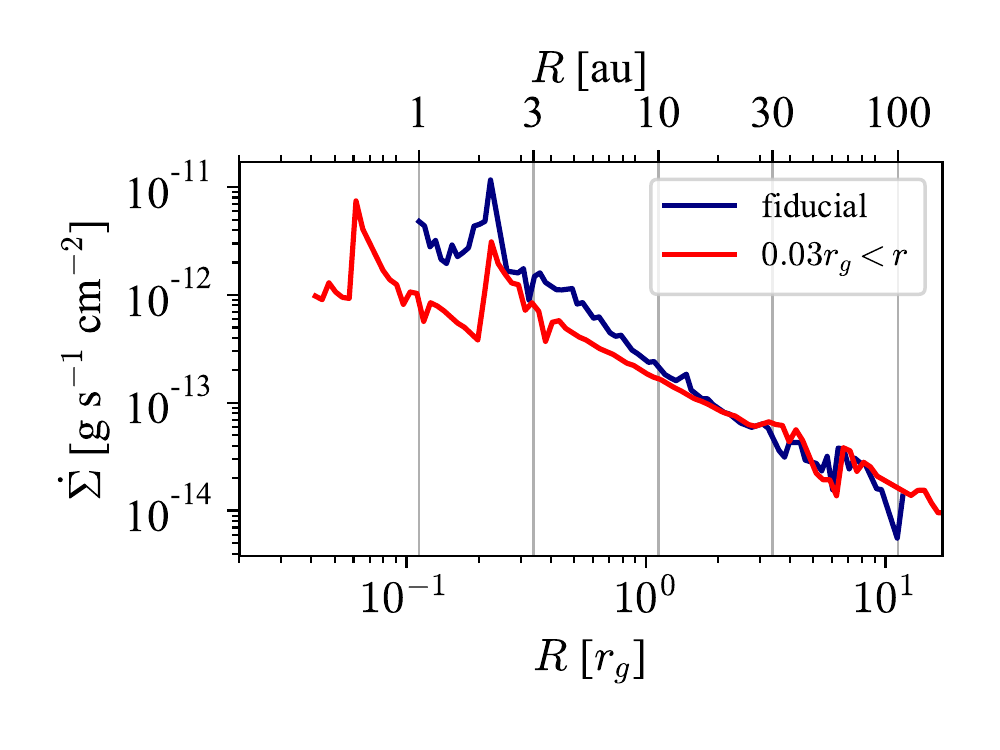}
         \caption{
         Comparison of the surface mass-loss rate between the simulations with computational domains of $0.03\rg \leq r \leq 20\rg$ (red) and $0.1\rg \leq r \leq 20 \rg$ (blue).}
      \label{fig:rmin003}
\end{figure}
We show the surface mass-loss rate in \figref{fig:rmin003}. 
We find that photoevaporation occurs also at $r < 0.1\rg$.
In the simulation with $0.03\rg\leq r\leq 20\rg$, we find a local maximum in the mass-loss rate profile at the same position ($R\sim0.17\rg$) as the fiducial one, but also another peak around $r\approx 0.06\rg$, which corresponds to the base density maximum.
In the simulation with the small inner boundary, the photoevaporative flows excited from $0.03\rg\leq R \leq 0.1\rg$ shield the stellar radiation that would have reached the farther outer region. Then, the height of the disk surface where $A_{\rm V} = 0.5$ increases compared to that in the fiducial run. This difference is significant in terms of density, 
and thus the surface mass-loss rate is {\it lower} in $0.2\rg<R<0.7\rg$ for the simulation with the small inner boundary.
We find that, although the disk structure at the innermost region is affected,
resolving the inner region does not significantly change the total mass-loss rate.

We calculate $2\pi R^{2}\dot{\Sigma}(R)$ based on $\dot{\Sigma}(R)$.
This value clarifies the mass loss by photoevaporation at each position.
\figref{fig:surfacedensity} shows how $2\pi R^{2}\dot{\Sigma}(R)$ differs by the distance from the central star.
The figure shows that the total mass-loss rates are dominated by the outer region. It results in the tendency that the total mass-loss rates increase with the radii at which they are measured.



\subsection{Semianalytic Model}\label{sec:result3}
\begin{figure}
         \includegraphics[width=\linewidth,clip]{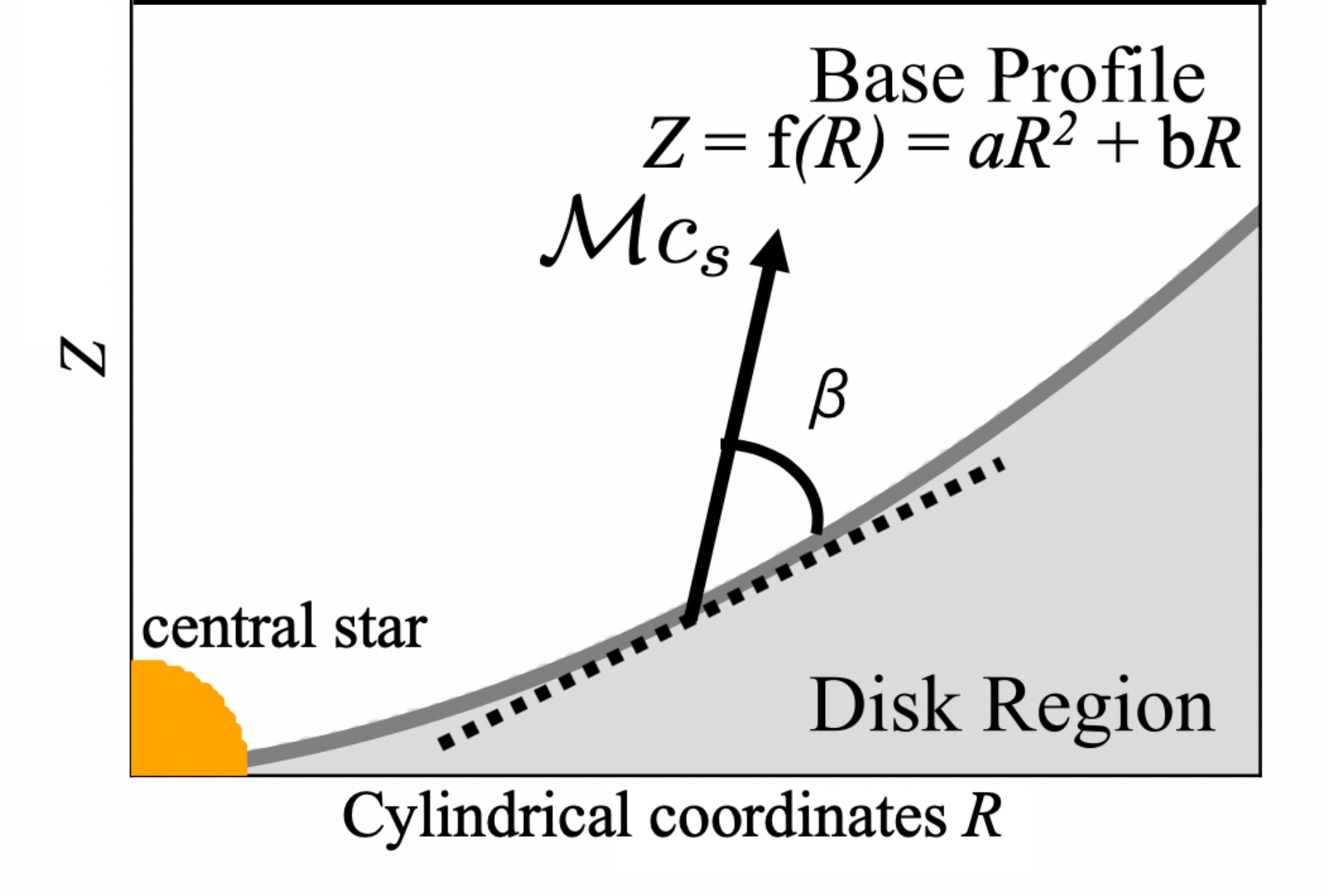}
         \caption{Configuration in our analytic model. We approximate the shape of the disk to be a quadratic function. The photoevaporative flows are launched at the speed $\mathcal{M}c_{s}$. We set the angle between the direction of the gas velocity and the base as $\beta$.}
      \label{fig:modelimage}
\end{figure}

We develop a semianalytic model to understand the physical
processes that cause the obtained stellar mass dependence of the mass-loss rates.
A schematic picture of our model is shown in \figref{fig:modelimage}.
In this model, the disk is in a steady state, and the photoevaporation flows are excited where the column density of \ce{H2} is $N(\ce{H2}) = 10^{20}{\,\rm cm}^{-2}$. This is the definition of the base in our model, and is determined empirically from the results of our simulations. The locus of the base is fitted by a function $Z = f(R, M_*) = a(M_*)R^{2} + b(M_*)R$ with the fitting coefficients $a, b$ being  functions of stellar mass $M_*$. The flows are launched from the base, having a Mach number of $\mathcal{M}$ and launching angle $\beta$ with respect to the base. The azimuthal velocity is simply assumed to be $v_\phi = \sqrt{G M_*/r}$ at the base. Following \eqnref{eq:massloss}, the launched photoevaporated flows are regarded as gravitationally unbound if $\eta > 0$. 

The coefficients for $f(R, M_*)$ are found to be
\[
\begin{split}
a(M_{*}) &= \biggl[3.81\times 10^{-5}\left(\frac{M_{*}}{\Msun}\right)^{2} - 5.44\times 10^{-3}\left(\frac{M_{*}}{\Msun}\right)\\
& + 4.58\times 10^{-2}\biggr]\times \rg^{-1}\\
b(M_{*}) &= 9.78\times 10^{-1}\left(\frac{M_{*}}{\Msun}\right) + 9.74\times 10^{-3}.
\end{split}
\]
We derive the base temperature of our model by directly fitting our simulation results as
\[
T_{\rm fit} = T_{0}(M_{*})\left(\frac{r}{\rg}\right)^{\alpha (M_{*})},
\]
where we find
\[
\begin{split}
T_{0} &= 1.30\times 10^{3} + 2.79\times 10^{2}\left(\frac{M_{*}}{\Msun}\right)\ \Kelvin\\
\alpha (M_*) &= 8.92\times 10^{-1} \left(\frac{M_{*}}{\Msun}\right)^{-0.108}.
\end{split}
\]
Similarly, we fit the base density of our simulations as 
\[
\rho_{\rm fit} = \rho_{0}(M_{*})\left(\frac{r}{\rg}\right)^{c(M_{*})},
\]
where
\[
\begin{split}
\rho_{0}(M_{*}) &= 3.03\times10^{-18}\left(\frac{M_{*}}{\Msun}\right)^{-0.309}\rm{\,g\, cm^{-3}}\\
c(M_{*})&= 3.60\times10^{-3}\left(\frac{M_{*}}{\Msun}\right)^{3}-2.55\times 10^{-2}\left(\frac{M_{*}}{\Msun}\right)^{2}\\
&-4.42\times 10^{-2}\left(\frac{M_{*}}{\Msun}\right)-0.640 .\\
\end{split}
\]
With the assumptions of our model, \eqnref{eq:enthalpy} reduces to
\[
\begin{split}
\eta = \frac{1}{2}\mathcal{M}^{2}c_{s}^{2}+\frac{\gamma}{\gamma -1}c_{s}^{2} - \frac{GM_{*}}{2r}
\end{split}
\]
at the base,
and the photoevaporation condition, $\eta > 0$, is rewritten as 
\[
r>r_{\rm min}
=\rg\left[ \frac{\gamma-1}{(\gamma-1)\mathcal{M}^{2}+2\gamma}\frac{\mu m_{\rm H}GM_{*}}{kT_{0}\rg} \right]^{\frac{1}{1+\alpha}}.
\]
This can be interpreted as the heated gas being unbound at $r > r_{\rm{min}}$. Note that $r_{\rm min}$ is defined in spherical polar coordinates. The corresponding cylindrical radius, $R_{\rm min}$, is given by the positive root of the equation, $r^2_{\rm min} = R_{\rm min}^2 + [f(R_{\rm min}, M_*)]^2$.
We find that $\mathcal{M}$ and $\beta$ are not strongly dependent on the distance from the central star.
The averaged $\mathcal{M}$ and $\beta$ are 
\[
\begin{gathered}
\bar{\mathcal{M}} = 0.35\times \left(\frac{M_{*}}{\Msun}\right)^{0.40},\\ 
\sin\bar{\beta} = 0.5,
\end{gathered}
\]
respectively. 
We adopt these values for our semianalytic model.

\begin{figure}
         \includegraphics[width=\linewidth,clip]{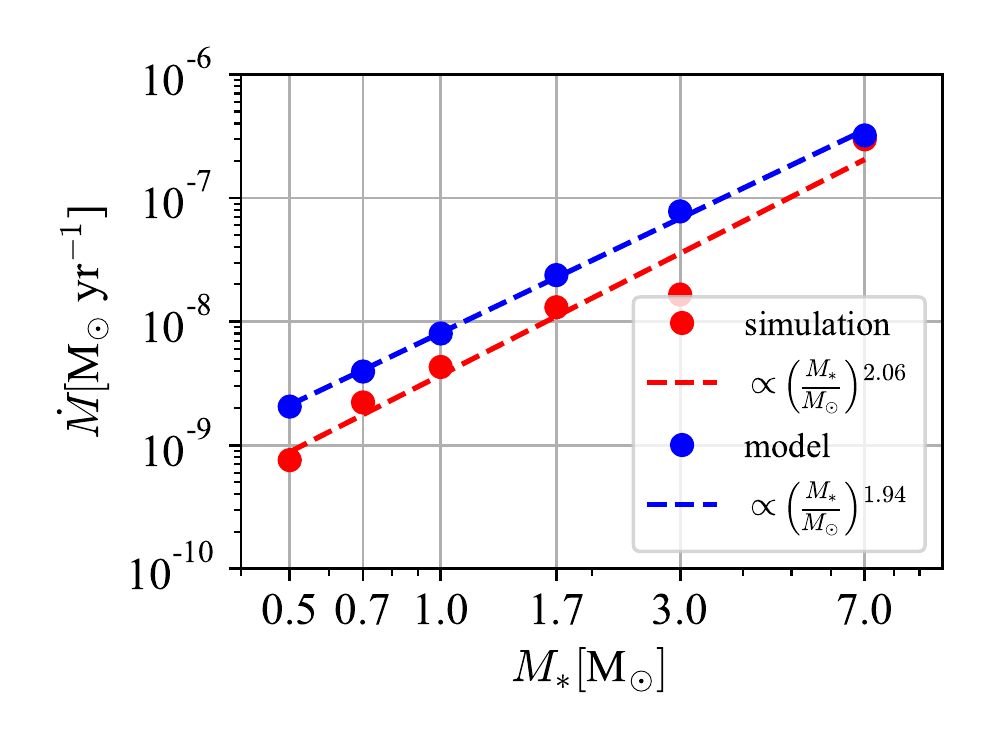}
         \caption{Mass-loss rates as a function of the host star mass. The red dots are the simulation results, and the blue dots shows the model photoevaporation rate. 
         The dashed lines with the corresponding colors show power-law fits.}
      \label{fig:modelmassloss}
\end{figure}

The cumulative photoevaporation rate within $R$ is computed by integrating the effective mass flux with respect to $R$,
\[
\dot{M}_{\rm{model}} = 2\times \int_{R_{\rm min}}^R\dd s\,2\pi R\, \rho v_{p}\,\sin\beta,
\]
where $\dd s$ is the line element of the base ($\dd s=\sqrt{1+{f^{\prime}}^{2}}\dd R$).
The gradient of $f$ is given by an average as  $\bar{f^{\prime}}=(f(R)-f(R_{\rm{min}}))/(R-R_{\rm{min}})$.
Substituting the fit for relevant quantities, we obtain
\begin{eqnarray}
&\dot{M}&_{\rm{model}} (< R) 
 = \frac{8\pi\bar{\mathcal{M}}\sin\bar{\beta}\rho_{0}\sqrt{{kT_{0}}/{\mu m_{\rm H}}}\sqrt{1+\bar{{f^{\prime}}^{2}}}}{\bar{({f^{\prime}}^{2}}+b\bar{f^{\prime}}+2)(2+c+\alpha/2)}\nonumber \\
 &\times& \rg^{2}\left(\frac{r_{\rm{max}}/\rg}{\sqrt{1+\bar{{f^{\prime}}^{2}}}}\right)^{2+c+\alpha/2} 
 \left[ 1-\left(\frac{r_{\rm{min}}}{r_{\rm{max}}}\right)^{2+c+\alpha/2} \right]\nonumber \\
 &\simeq& 2.97 \times 10^{-10} \Msun\yr^{-1}\,\left(\frac{M_{*}}{\ce{M}_{\odot}}\right)^{2.40} 
 \left(\frac{r_{\rm max}/\rg}{\sqrt{1+\bar{{f^{\prime}}^{2}}}}\right)^{2+c+\alpha/2}
\nonumber \\
&\times& \rho_{18} T_{3}^{1/2} 
\frac{\sqrt{1+\bar{{f^{\prime}}^{2}}} 
\left[ 1-\left({r_{\rm{min}}}/{r_{\rm{max}}}\right)^{2+c+\alpha/2} \right]}{(\bar{{f^{\prime}}^{2}}+b\bar{f^{\prime}}+2)(2+c+\alpha/2)},\,\,\,\,\, \nonumber
\end{eqnarray}
where $r_{\rm max} \equiv \sqrt{R^2 + [f(R^2, M_*)]^2}$, 
$\rho_{18} \equiv \rho_0/ 10^{-18}{\rm \,g\,cm^{-3}}$,
and $T_{3} \equiv T_0/10^3\Kelvin$.
In \figref{fig:modelmassloss}, we compare the analytic mass-loss rates within $r < 20\rg$ with those of the simulations.

The semianalytic model can accurately reproduce the simulation results. Note that the contribution of the mass flux within $< r_{\rm min}$ is 1--4\% of the total mass-loss rates in our simulations. Therefore, the mass-loss rates of the model and simulations are fairly in good agreement despite the fact that the model ignores the contribution of the hot gas from $< r_{\rm min}$.

\subsection{Luminosity Dependence}\label{sec:atsomepoint}
The FUV, EUV, and X-ray luminosities of young stars are 
not tightly constrained by observations, mainly because of  uncertain interstellar extinction.
The luminosities are likely different among the same spectral-type stars, and can also depend on other stellar properties such as age, magnetic activities, and accretion rate \citep[e.g.,][; see also \cite{Alexander:2014} for review]{Gullbring:1998, 2005Preibisch, 2011Ingleby, Gudel:2007, Yang:2012, 2012:Schindhelm, France:2012, Vidotto:2014, France:2018}. 
In order to examine the effects of the assumed luminosities on our results, we perform additional simulations for $M_* = 1\Msun$ but with varying each of FUV, EUV, and X-ray luminosities by 0.01, 0.1, 10, and 100~times of the fiducial value.
The resulting mass-loss rates are $7.5$--$9.0\times 10^{-9}\Msun\yr^{-1}$ when varying the EUV luminosity. 
Clearly, the mass-loss rate is insensitive to the EUV luminosity. 
\begin{figure}
         \includegraphics[width=\linewidth,clip]{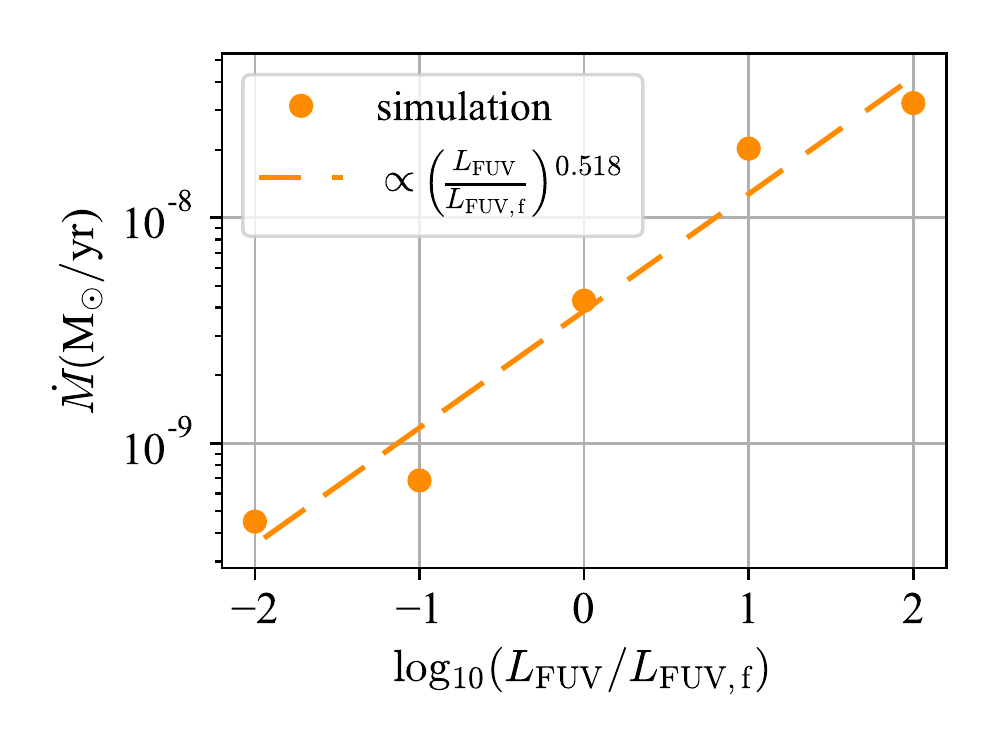}
         \includegraphics[width=\linewidth,clip]{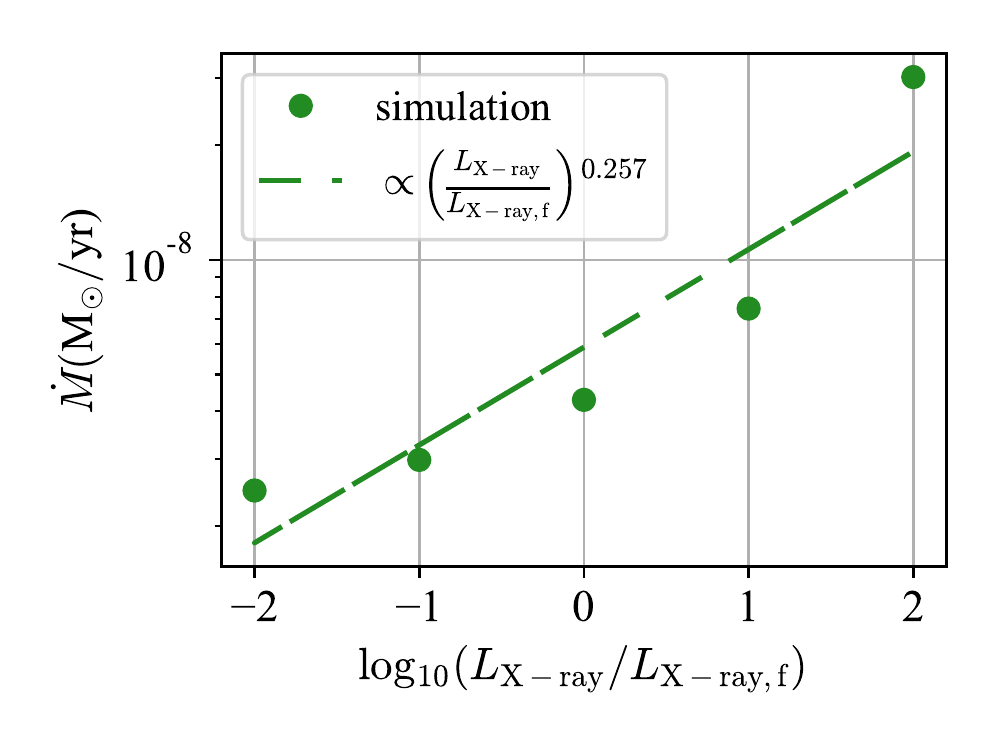}
         \caption{(top) Mass-loss rate with each FUV luminosity and $M_{*}=1.0\Msun$. The orange points are the simulation data. The orange dashed line is a fit. 
         (bottom) Mass-loss rate with each X-ray luminosity and $M_{*}=1.0\Msun$.
         The green points are the simulation data. 
         The green dashed line is a fit.
         $L_{\rm{f}}$ is a fiducial value defined in \tabref{table:stellarparameter}
         }
      \label{fig:FLXL}
\end{figure}
In contrast, the mass-loss rate increases with increasing FUV and X-ray luminosities as shown in \figref{fig:FLXL}.
We find an approximate dependence of

\begin{gather}
    \dot{M}\simeq3.86\times10^{-9}\left(\frac{L_{\rm{FUV}}}{L_{\rm{FUV},\rm{f}}}\right)^{0.518}\,\Msun\yr^{-1}    \label{eq:fuvdependence}\\
    \dot{M}\simeq 5.90\times 10^{-9}\left(\frac{L_{\rm{X-ray}}}{L_{\rm{X-ray},\rm{f}}}\right)^{0.257}\,\Msun\yr^{-1},
\end{gather}
where $L_{\rm{FUV},\rm{f}}$ and $L_{\rm{X-ray},\rm{f}}$ are the fiducial values (Table 1).
X-ray is responsible for the mass-loss rate for $L_{\rm FUV}/L_{\rm FUV, f} = 0.01$ and $L_{\rm X-ray}/L_{\rm X-ray, f} \gtrsim 10$. 
In the Appendix~\ref{app:sigmadot}, we provide fits to the surface mass-loss rates of the simulations with various luminosities. 
The FUV and X-ray heating rates are comparable when $L_{\rm FUV}/L_{\rm FUV,f} \approx 0.05$ ($L_{\rm FUV}/L_{\rm X} \sim 1$). The higher $L_{\rm FUV}$/$L_{\rm X}$ ratio in our
fiducial runs promotes FUV heating that dominates over X-ray heating (see \figref{fig:heating}). 

It is worth comparing our results with those of previous studies.
\cite{2019Picogna} run hydrodynamics simulations with varying X-ray luminosities. They find that the mass-loss rate scales with $L_{\rm X-ray}$ more steeply
than our result (\figref{fig:FLXL}).
In our model, photoevaporation is predominantly driven by FUV radiation when the X-ray luminosity is low, 
and thus the mass-loss rate does not scale strongly with the X-ray luminosity if $L_{\rm X-ray}/L_{\rm X-ray, f} \lesssim 10$.
This leads to apparently different X-ray luminosity dependence from that of \cite{2019Picogna}.
On the other hand, with strong X-ray radiation,
the disk heating is dominated by X-ray, and the mass-loss rate increases as $\dot{M} \propto (L_{\rm X-ray}/L_{\rm X-ray, f})^{0.6}$, which is 
close to the result of \cite{2019Picogna}.

Since the FUV luminosity decreases as the accretion rate decreases, X-ray heating would become important at later phases of the disk evolution.
Note, however, that there are several other effects that can affect the relative dominance of the FUV/EUV/X-ray heating. For instance, grain growth and the stellar evolution
itself can affect the heating efficiencies in a complicated manner.  
Including those effects would be necessary to fully understand the chronological variability of the disk structure and the mass-loss rate. 



\section{Discussion}
\subsection{Dispersal Time of Inner Disks}\label{sec:lifetime}
Previous 1D disk evolution models suggest that the surface density evolution is governed by viscous evolution at an early stage. When the accretion rate decreases comparable to the photoevaporation rate, a gap opens at around the gravitational radius, which separates the initially smooth disk into the inner and outer disks \citep[e.g.,][]{Clarke:2001, Alexander:2006, Owen:2010}. The inner disk quickly accretes onto the central star on the viscous timescale, while the outer disk is dispersed by photoevaporation on a relatively short time compared to the disk lifetime. In this section, we estimate the inner-disk lifetimes using the photoevaporation rates we have derived so far. We compare them with those estimated by infrared observations. 

We define the inner-disk lifetime as the time at which the accretion rate decreases to the photoevaporation rate.
Following \citet{Clarke:2001}, we introduce the viscous scaling time
\[
    t_{\rm s} = \frac{M_{\rm{disk, 0}}}{2\dot{M}_{\rm{acc, 0}}}, 
\]
where $M_{\rm disk, 0}$ and $\dot{M}_{\rm acc,0}$ are the initial disk mass and the initial rate of accretion onto the star, respectively. The accretion rate decays as $\propto t^{-3/2}$ for $t \gg t_{\rm s}$. Thus, the timescale $t_{\rm w}$ over which $\dot{M}_{\rm acc}$ decreases to $\dot{M}$ can be derived by solving 
\begin{equation}
\dot{M} \approx \dot{M}_{\rm acc,0} \braket{\frac{t}{t_{\rm s}}}^{-3/2} \label{eq:tw}
\end{equation}
with respect to $t$ \citep{Clarke:2001}. 
In our fiducial model, $\dot{M}$ is determined by FUV-driven photoevaporative flows. 
The adopted accretion rate has is $\dot{M}_{\rm acc} = 3\times 10^{-8} (M_*/\Msun)^2 \Msun \yr^{-1}$ \citep{Muzerolle:2003, GortiHollenbach:2009} which is assumed to be that of a $1\Myr$-old system \citep{GortiHollenbach:2009}. The accretion rate is supposed to be significantly higher at an earlier stage of the disk evolution. The FUV luminosity and thereby $\dot{M}$ are also expected to be higher at that time and to be time dependent. To incorporate this effect into our discussion here, we assume $L_{\rm FUV} \propto \dot{M}_{\rm{acc}}$ \citep[e.g.,][]{1998_CalvetGullbring, Gullbring:1998} and $\dot{M} \propto L_{\rm FUV} ^{\lambda}$ ($0 < \lambda < 1$; see \eqnref{eq:fuvdependence}). These assumptions reduce \eqnref{eq:tw} to 
\[
    t_{\rm w} = t_{\rm s} \left[\frac{\dot{M}_{\rm acc, 0}}{\dot{M}(1\Myr)} \braket{\frac{\dot{M}_{\rm acc} (1\Myr)}{\dot{M}_{\rm acc, 0}}}^\lambda\right]^{{2}/{3(1-\lambda)}}, 
\]
where $\dot{M}_{\rm acc} (1\Myr) = 3\times 10^{-8} (M_*/\Msun)^2 \Msun \yr^{-1}$, and $\dot{M} (1\Myr)$ is the photoevaporation rate obtained in this study. Since $t_{\rm w}$ is determined for given $M_{\rm disk, 0}$ and $\dot{M}_{\rm acc,0}$, we treat the latter two quantities as parameters in the following discussion. We adopt the power-law index of \eqnref{eq:fuvdependence} for $\lambda$.

\begin{figure}
         \includegraphics[width=\linewidth,clip]{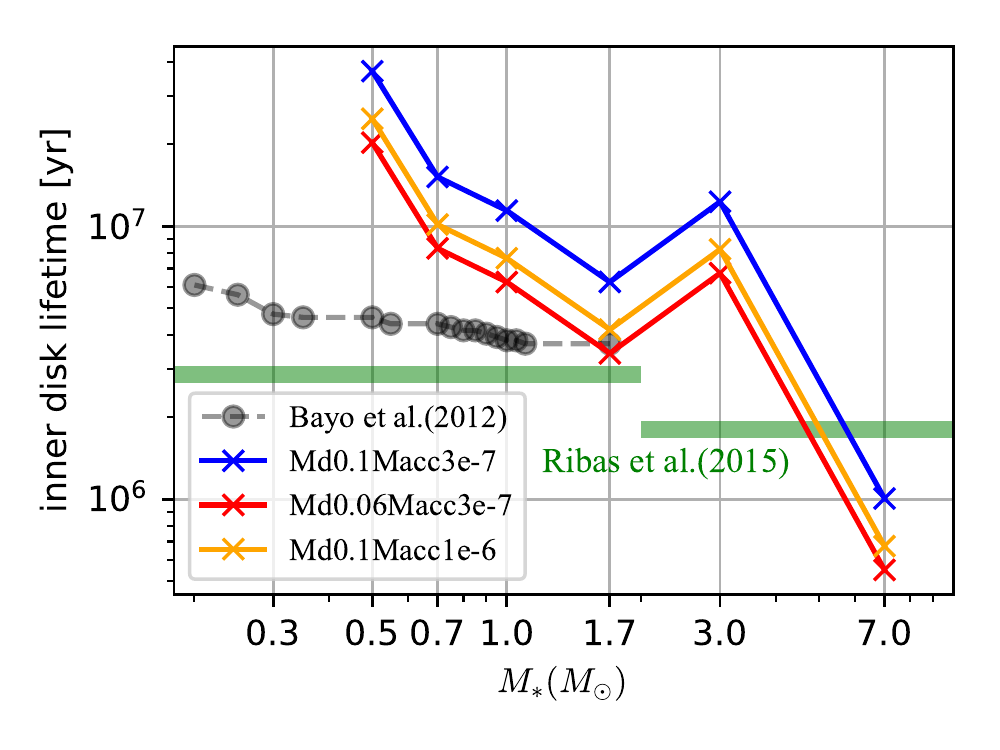}
         \caption{
         Inner-disk lifetimes. The blue, red, and orange solid lines are those derived with assuming the initial disk mass and the accretion rate onto the star as $(M_{\rm disk,0}, \dot{M}_{\rm acc,0}) = (0.1\,M_*, 3.0\times10^{-7}\Msun\yr^{-1}), (0.06\,M_*, 3.0\times10^{-7}\Msun\yr^{-1}), (0.1\,M_*, 1.0\times10^{-6}\Msun\yr^{-1})$, respectively. 
          The black points and green horizontal bars represent inner-disk lifetimes estimated from observational disk fractions in \cite{2012Bayo} and \cite{Ribas:2015}, respectively. See the main text for more detail. }
      \label{fig:AllLifetime}
\end{figure}

The resulting $t_{\rm w}$ are shown in \figref{fig:AllLifetime}. We have considered three parameter sets: 
\[
(M_{\rm disk,0}, \dot{M}_{\rm acc,0}) = \left\{ 
\begin{array}{l}
    (0.1\,M_*, 3.0\times10^{-7}\Msun\yr^{-1})  \\
     (0.06\,M_*, 3.0\times10^{-7}\Msun\yr^{-1}) \\
     (0.1\,M_*, 1.0\times10^{-6}\Msun\yr^{-1})
\end{array}
\right. .
\] 
These models are labeled as ``Md0.1Macc3e-7'', ``Md0.06Macc3e-7'', and ``Md0.1Macc3e-6'' in \figref{fig:AllLifetime}, and are shown as blue, red, and orange circles, respectively. 
The derived $t_{\rm w}$, our proxy for the inner-disk lifetime, is shorter for the models that yield a higher mass-loss rate (\figref{fig:simulationfuvmassloss}). 

In the discussion above, we have followed \cite{Clarke:2001} and have assumed a viscosity profile of $\nu \propto R$.
We can also consider different radial profiles as $\nu \propto R^{\kappa}$, to examine the uncertainty in the derived lifetimes.
To this end, we adopt a generalized model of \cite{1998Hartmann} and rewrite \eqnref{eq:tw} as
\[
\dot{M} \approx \dot{M}_{\rm acc,0} \braket{\frac{t}{t_{\rm s}}}^{-(5/2-\kappa)/(2-\kappa)}.
\]
Then, the inner-disk lifetime is expressed as
\[
\begin{split}
    t_{\rm w} &= t_{\rm s} \left[\frac{\dot{M}_{\rm acc, 0}}{\dot{M}(1\Myr)} \braket{\frac{\dot{M}_{\rm acc} (1\Myr)}{\dot{M}_{\rm acc, 0}}}^\lambda\right]^{\chi}\\
    &\chi = \frac{(2-\kappa)}{(5/2-\kappa)(1-\lambda)}\ .
\end{split}
\]
If we adopt $\kappa=3/2$ as in one of the variable
$\alpha$-viscosity model of \cite{1998Hartmann}, 
the resulting lifetime becomes twice as short as the fiducial value, but the stellar mass dependence remains essentially unaffected.

In \figref{fig:AllLifetime}, we also plot the inner-disk lifetimes estimated from the observational disk fractions compiled in \cite{2012Bayo} and \cite{Ribas:2015}. 
\cite{2012Bayo} provide a mass function of the disk fractions obtained by infrared excess observations toward the young cluster Collinder~69 in the Lambda Orionis star-forming region. The disk fraction at a mass bin $M_i$ is defined as the ratio of the disk-bearing members to the total members in $M_* < M_i$. Although this definition does not give the specific disk fraction at $M_* = M_i$, we use the disk fractions of \cite{2012Bayo} as an approximate substitute of the specific disk fraction; otherwise, the number of samples is too small to be statistically meaningful in each mass bin.
We convert the disk fraction to lifetime in an approximate manner as 
\[
t_{\rm life} = - \frac{t_{\rm age}}{\ln f_{\rm disk}},
\]
where $f_{\rm disk}$ is the disk fraction and $t_{\rm age}$ is the age of the star-forming region \citep[$t_{\rm age} = 5\Myr$;][]{Bayo:2011}. 
The resulting lifetimes are shown by the gray circles in \figref{fig:AllLifetime}. 
\cite{Ribas:2015} derive the disk fractions using a large sample across various star-forming regions whose ages are in the range of $\sim 1$--$11\Myr$. The samples are categorized into low-mass ($M_* < 2\Msun$) and high-mass ($M_* > 2\Msun$) sources. Each group is further classified into young ($1$--$3\Myr$) and old ($3$--$11\Myr$) sources, and the disk fractions are derived for each of the four categories. We approximately estimate the lifetimes of the low-mass and high-mass sources by fitting the disk fractions as a function of age 
\[
 f (t) = \exp\left[-\frac{t}{t_{\rm life}} \right]. 
\]
Each of the age bins has a wide spread. Therefore, we take the medians, $2\Myr$ and $7\Myr$ for the young and old groups, respectively, as representative ages. The resulting lifetimes are represented by the green horizontal bars in \figref{fig:AllLifetime}.

The stellar mass dependence of the estimated inner-disk lifetimes is overall consistent with the observations. The inner-disk lifetimes are estimated to be much longer than the observations for $M_* \leq 1.7 \Msun$. This may indicate that other processes, such as MHD winds and planet formation, contribute to dispersing the disks in this stellar mass range. Interestingly, recent 1D simulations that implement all of accretion, photoevaporation, and MHD winds suggest that MHD winds drive a major mass loss at the early epoch of the evolution at least for solar-type stars \citep[][Fukuhara in prep.]{Kunitomo:2020}. We also find that the inner-disk lifetimes are long for $M_* \sim 3\Msun$. This suggests that the stellar (evolutionary) property can cause a non-monotonic dependence of the lifetimes on the stellar mass. The estimated lifetimes are significantly short at $M_* > 3\Msun$ because the strong photoevaporation disperses the disk quickly around a high-mass star. 


\subsection{Uncertainties of Model Parameters}

The abundance of PAHs and small grains is critical for FUV photoelectric heating and hence FUV-driven photoevaporaiton \citep{Gorti:2008, GortiHollenbach:2009, Gorti:2015, Nakatani:2018a}. The dust abundance can vary significantly 
through grain growth and destruction by the stellar UV and X-ray radiation \citep{Siebenmorgen:2010, Vicente:2013}. Observationally, the PAH abundance is poorly constrained, but it has been suggested to be smaller than the interstellar value \citep{Geers:2007, Oliveira:2010}. 
\cite{Gorti:2015} show that dust growth indeed results in weakening FUV photoevaporation, but FUV radiation is still effective enough to drive mass loss with the reduced PAH and small grain abundances by a factor of $\sim10$--$100$. 
This is also expected in our model, and thus the derived dispersal time would not be strongly affected by taking into account reduced abundances of PAHs and small grains.
Note that, for further lower dust abundances, X-ray heating is the
primary process that drives mass loss. 


\subsection{Other Effects}
\figref{fig:AllLifetime} shows a stellar mass dependence of dispersal time caused by {\it internal} photoevaporation. 
Photoevaporation by external radiation sources has been suggested to shape the mass functions of disk masses in star-forming regions by recent submillimeter surveys \citep{Ansdell:2018}. Distance from the external UV source(s) and the disk size (the geometrical cross section to the radiation field) are key parameters to determine the mass-loss rate in such cases \citep{JohnstoneHollenbachBally:1998, StorzerHollenbach:1999, HollenbachYorkeJohnstone:2000}. 
It would be important to incorporate the effects of external photoevaporation to address the disk lifetime quantitatively, particularly for massive star-forming regions. 
The derived mass-loss rates in this study apply to the systems where internal photoevaporation dominates mass loss.

We have derived the disk-dispersal time by a scaling argument
on the basis of physical processes. In practice, accretion and MHD winds can also contribute considerably to the mass evolution. These effects are responsible for angular momentum redistribution and dominate mass loss especially at an early evolutionary stage of protoplanetary disks \citep{Suzuki:2016, Kunitomo:2020}. 
As discussed in \secref{sec:lifetime}, disk accretion is also an important process to determine the overall impact of FUV-driven photoevaporation. When the system evolves and the accretion rate drops, the photoevaporation rate can also decrease owing to the reduced
FUV contribution from the accretion. Furthermore, the stellar FUV/EUV/X-ray luminosities can vary depending on detailed processes and phases in the stellar evolution such as development of the convective zone and changes in the photospheric temperature. Incorporating these variable processes over the evolution of the system would also significantly affect the disk-dispersal process and the stellar mass dependence. 

For more detailed comparisons between our model and observations, it may be necessary to follow a long-term evolution including a variety of physical processes.

\section{summary}
Recent observations suggest that the lifetime of a PPD depends on the central stellar mass.
Theoretically, disk photoevaporation can cause such  dependence because the luminosity of the central star
increases with the stellar mass. On the other hand, 
stronger gravity of heavier stars may prevent efficient 
disk mass loss, and the details of the disk-dispersal mechanism remain unknown.
We have performed a suite of radiation hydrodynamics simulations to investigate the disk lifetime.
Our simulations show that the mass-loss rate 
increases with $M_*$ as $\dot{M} = 3.69\times10^{-9}(M_{*}/\Msun)^{2.06}\Msun\yr^{-1}$.

 Based on the simulation results, we have developed a semianalytic model that reproduces the disk mass-loss rates
 accurately.
We derive the surface mass-loss rates
for a variety of cases and provide simple polynomial functional fits. The results are aimed at being used 
by semianalytic studies of the evolution of PPDs.
We have found that there is a nonzero contribution to the mass-loss rate even within $r<\rg$.
We have estimated the inner-disk lifetime from the resulting photoevaporation rate.
The estimated inner-disk lifetime is consistent with the proposed observational trend of the stellar mass dependence, where the lifetimes are shorter for higher-mass stars. 

Our radiation hydrodynamics simulations show that disk photoevaporation can explain the observed correlation between disk lifetime and the host star mass. 
We have not included the effects of accretion, MHD winds, and external radiation sources in the present study. 
In future work, we will study the long-term evolution of PPDs including these effects and make a comparison with observations.



\acknowledgments
We thank Takeru Suzuki and Masahiro Ogihara for fruitful discussions. 
R.N. acknowledges support from the Special Postdoctoral Researcher program at RIKEN and from Grant-in-Aid for Research Activity Start-up (19K23469).
Numerical computations were carried out on Cray XC50 at Center for Computational Astrophysics, National Astronomical Observatory of Japan
%
\vspace{5mm}




\appendix

\section{Mass loss profiles for various input luminosities} \label{app:sigmadot}
We calculate the surface mass-loss rates for various FUV and X-ray luminosities considered in \secref{sec:atsomepoint}.
We fit $\dot{\Sigma}(R)$ by
\[
\begin{split}
\log_{10} \braket{\frac{\dot{\Sigma}(R)}{1\,{\rm g\,s^{-1}\,cm^{-2}}}} &= c_8\,x^8+c_7\,x^7 +c_6\,x^6\\
&+c_5\,x^5 +c_4\,x^4+c_3\,x^3\\
&+c_2\,x^2 +c_1\,x +c_0
\end{split}
\]
where 
\[
x  = \log_{10}\braket{\frac{R}{\rg}}.
\]
Here we resort to higher-order polynomial fitting than that of \eqref{eq:surfacemasslossrate} because we have found that \eqnref{eq:surfacemasslossrate} does not provide accurate fits to $\dot{\Sigma}(R)$. 
We list the obtained fitting coefficients ($c_0,c_1,c_2,c_3,c_4,c_5,c_6,c_7$, and $c_8$) in \tabref{tab:fit2}.

\begin{table*}
  \caption{The coefficients of $\dot{\Sigma}(R)$}
  \label{tab:fit2}
  \begin{tabular}{L|C C C C C C C C C}\hline
    $\rm{Luminosity}$ & c_8 & c_7 & c_6 & c_5 & c_4 & c_3 & c_2 & c_1 & c_0 \\ \hline
    $L_{\rm{FUV}, \rm{f}}\times0.01$ & -6.90 & 6.04 & 9.87 & -7.81 & -3.22 & 2.46 & -0.124 & -1.10 & -13.3 \\
    $L_{\rm{FUV}, \rm{f}}\times0.1$ & -12.7 & 10.4 & 18.0 & -14.0 & -6.32 & 5.80 & 0.347 & -1.90 & -13.2 \\
    $L_{\rm{FUV}, \rm{f}}\times10$ & -2.91 & 4.62 & 3.02 & -6.82 & 0.325 & 2.49 & -0.732 & -1.16 & -11.7 \\
    $L_{\rm{FUV}, \rm{f}}\times100$ & -14.1 & 21.5 & 11.3 & -24.6 & -1.19 & 8.33 & -0.273 & -2.15 & -11.2 \\
    $L_{\rm{X-ray}, \rm{f}}\times0.01$ & -7.36 & 6.47 & 8.34 & -6.88 & -1.68 & 1.87 & -0.32 & -1.18 & -12.6 \\
    $L_{\rm{X-ray}, \rm{f}}\times0.1$ & -3.50 & -0.029 & 7.08 & -0.422 & -3.67 & 0.695 & 0.166 & -1.34 & -12.4\\
    $L_{\rm{X-ray}, \rm{f}}\times10$ & 6.67 & -3.94 & -15.0 & 7.69 & 8.43 & -2.95 & -1.14 & -1.31 & -12.0\\
    $L_{\rm{X-ray, \rm{f}}}\times100$ & -5.43 & -2.02 & 10.0 & 1.34 & -5.95 & 0.637 & 1.31 & -1.38 & -11.5\\ \hline
  \end{tabular}
\end{table*}





\bibliography{bibliography}
\bibliographystyle{aasjournal}



\end{document}